\documentclass[10pt]{article}

\usepackage[top=1in, bottom=1in, left=1in, right=1in]{geometry}

\usepackage{setspace}
\usepackage{algorithm}

\usepackage{algpseudocode}

\usepackage{amssymb}

\usepackage{amsthm}

\usepackage{amsmath}

\usepackage{adjustbox}

\usepackage{bm}

\usepackage{bbm}

\usepackage{natbib}

\usepackage{graphicx}

\usepackage{booktabs}

\usepackage{blkarray}

\usepackage{titlesec}  

\usepackage{authblk}
\usepackage[labelfont=bf, textfont=it]{caption}

\DeclareMathOperator*{\argmin}{argmin}   %
\newcommand{\mat}[1]{\bm{#1}}

\newcommand{\vect}[1]{\bm{#1}}

\newcommand{\T}{\mathsf{T}}

\newtheorem{lemma}{Lemma}
\newtheorem{theorem}{Theorem}

\theoremstyle{definition}

\newtheorem{assumption}{Assumption}

\title{Data-fusion using factor analysis and low-rank matrix completion}

\author[1,*]{Daniel Ahfock}
\author[2,3]{Saumyadipta Pyne}
\author[1]{Geoffrey J. McLachlan}
\affil[1]{School of Mathematics and Physics, University of Queensland}
\affil[2]{ Public Health Dynamics Laboratory and Department of Biostatistics,\\ Graduate School of Public Health, University of Pittsburgh, PA, USA}
\affil[3]{Health Analytics Network, PA, USA}
\affil[*]{\texttt{d.ahfock@uq.edu.au}}
\date{}                     
\begin{document}

\maketitle

\begin{abstract}
Data-fusion involves the integration of multiple related datasets. The statistical file-matching problem is a canonical data-fusion problem in multivariate analysis, where the objective is to characterise the joint distribution of a set of variables when only strict subsets of marginal distributions have been observed. Estimation of the covariance matrix of the full set of variables is challenging given the missing-data pattern. Factor analysis models use lower-dimensional latent variables in the data-generating process, and this introduces low-rank components in the complete-data matrix and the population covariance matrix. The low-rank structure of the factor analysis model can be exploited to estimate the full covariance matrix from incomplete data via low-rank matrix completion.  We prove the identifiability of the factor analysis model in the statistical file-matching problem under conditions on the number of factors and the number of shared variables over the observed marginal subsets. Additionally, we provide an EM algorithm for parameter estimation.  On several real datasets, the factor model gives smaller reconstruction errors in file-matching problems than the common approaches for low-rank matrix completion. 
\end{abstract}

\section{Introduction}
Data-fusion involves the joint modelling of multiple related datasets, with the aim to lift the quality of inference that could be obtained from isolated separate analyses of each dataset. The statistical file-matching problem is a classic data-fusion task, where all observations are from the same population, with the caveat that different sets of variables are recorded in each individual dataset \citep{little_2002_statistical, rassler_2002_statistical, dorazio_2006_statistical}. The core problem can be reduced to the analysis of two datasets, labelled A and B, with three groups of variables denoted $\vect{X}, \vect{Y}$, and $\vect{Z}$. Dataset A contains observations on the $(\vect{X},\vect{Y})$ variables and Dataset B contains observations on the $(\vect{X}, \vect{Z})$ variables. Table \ref{tab:missing} describes the missing-data pattern. The statistical file-matching problem occurs in bioinformatics such as in flow cytomtery analysis in the synthesis of multiplexed data collected using different panels of markers \citep{pedreira_2008_generation, lee2011statistical, oneill_2015_deep, abdelaal_2019_cytofmerge}. Other important application areas include survey sampling and data integration in official statistics \citep{conti_2016_statistical, dorazio_2019_statistical}.

\begin{table}[h]
\begin{center}
\caption{Missing data pattern in the statistical file-matching problem}
\label{tab:missing}
\begin{tabular}{lccc}
\hline
 Variables & \multicolumn{1}{c}{$X$} &  \multicolumn{1}{c}{$Y$} & 
\multicolumn{1}{c}{$Z$} \\ \hline
Dataset A & observed & observed & --- \\
Dataset B & observed & --- & observed \\
\hline
\end{tabular}
\end{center}
\end{table}

We assume that dataset A and dataset B are samples from the same homogeneous population. For example, in flow cytometry analysis a single blood sample from a patient may be divided into two aliquots which are then analysed using markers $(\vect{X}, \vect{Y})$ and $(\vect{X}, \vect{Z})$ respectively. Technological limitations prevent joint measurements on the $\vect{Y}$ and $\vect{Z}$ markers. In survey sampling, one random subset of the population may receive a questionnaire with items $(\vect{X}, \vect{Y})$ and a second random subset of the population given items $(\vect{X}, \vect{Z})$. This situation may arise when integrating results from related surveys, or in the design of a single survey where there is a constraint on the response burden. 

Key linear relationships between $\vect{X}, \vect{Y}$, and $\vect{Z}$ are encoded in the covariance matrix $\mat{\Sigma}$:
\begin{align}
    \text{cov}\begin{pmatrix} \vect{X} \\ \vect{Y} \\ \vect{Z} \end{pmatrix} &=\mat{\Sigma}=\begin{pmatrix}
{\mat{\Sigma}}_{XX} & {\mat{\Sigma}}_{XY} & {\mat{\Sigma}}_{XZ} \\
{\mat{\Sigma}}_{YX} & {\mat{\Sigma}}_{YY} & {\mat{\Sigma}}_{YZ} \\
{\mat{\Sigma}}_{ZX} &  {\mat{\Sigma}}_{ZY} &{\mat{\Sigma}}_{ZZ}
\end{pmatrix}.
\end{align}
A fundamental task in the statistical file-matching problem is the estimation of $\mat{\Sigma}_{YZ}$ from incomplete data \citep{rassler_2002_statistical, dorazio_2006_statistical}.  A common objective in file-matching is to impute the missing observations so that complete-data techniques can be used in downstream tasks. Recovery of $\mat{\Sigma}_{YZ}$ is crucial to generating proper imputations \citep{little_1988_missing, van_2018_flexible}.  The covariance matrix is also important for Gaussian graphical modelling, and the recovery of a network model from incomplete information is an interesting problem \citep{sachs_2009_learning}. 

The critical issue in the statistical file-matching problem is how to estimate $\mat{\mat{\Sigma}}_{YZ}$ without any joint observations on the $\vect{Y}$ and $\vect{Z}$ variables. Factor analysis models are useful in data-fusion tasks as they provide structured covariance matrices that allow the sharing of information across variables \citep{li_2017_incorporating, oconnell_2019_linked, park_2020_integrative}. In the statistical file-matching problem, this can facilitate the estimation of $\mat{\mat{\Sigma}}_{YZ}$ from the observed $(\vect{X}, \vect{Y})$ and $(\vect{X}, \vect{Z})$ associations. Let $p_{X}, p_{Y}$, and $p_{Z}$ denote the dimensions of $\vect{X}$, $\vect{Y}$, and $\vect{Z}$ respectively, with $p=p_{X}+p_{Y}+p_{Z}$. A factor analysis model consists of a $p \times q$ matrix of factor loadings $\mat{\Lambda}$ and a diagonal $p \times p$ matrix $\mat{\Psi}$ of uniqueness terms, giving the structured covariance matrix $\mat{\mat{\Sigma}}= \mat{\Lambda} \mat{\Lambda}^{\T} + \mat{\Psi}$. The full covariance matrix $\mat{\mat{\Sigma}}$ can be written as
\begin{align}
    \mat{\mat{\Sigma}} &= \mat{\Lambda} \mat{\Lambda}^{\T} + \mat{\Psi} = \begin{pmatrix}
    \mat{\Lambda}_{X}\mat{\Lambda}_{X}^{\T} + \mat{\Psi}_{X} & \mat{\Lambda}_{X}\mat{\Lambda}_{Y}^{\T} & \mat{\Lambda}_{X}\mat{\Lambda}_{Z}^{\T} \\
    \mat{\Lambda}_{Y}\mat{\Lambda}_{X}^{\T} & \mat{\Lambda}_{Y}\mat{\Lambda}_{Y}^{\T} + \mat{\Psi}_{Y} & \mat{\Lambda}_{Y}\mat{\Lambda}_{Z}^{\T} \\
    \mat{\Lambda}_{Z}\mat{\Lambda}_{X}^{\T} & \mat{\Lambda}_{Z}\mat{\Lambda}_{Y}^{\T} & \mat{\Lambda}_{Z}\mat{\Lambda}_{Z}^{\T} + \mat{\Psi}_{Z}
    \end{pmatrix}, \label{eq:sigma}
\end{align}
where $\mat{\Lambda}$ and $\mat{\Psi}$ have been partitioned as
\begin{align*}
    \mat{\Lambda} = \begin{pmatrix}
    \mat{\Lambda}_{X} \\
    \mat{\Lambda}_{Y} \\
    \mat{\Lambda}_{Z}
    \end{pmatrix}, \quad  \mat{\Psi} = \begin{pmatrix}
    \mat{\Psi}_{X} & \mat{0} & \mat{0} \\
    \mat{0} & \mat{\Psi}_{Y} & \mat{0} \\
    \mat{0} & \mat{0} & \mat{\Psi}_{Z}
    \end{pmatrix},
\end{align*}
where $\mat{\Lambda}_{X}$ is a $p_{X} \times q$ matrix of factor loadings for the $\vect{X}$ variables, and $\mat{\Psi}_{X}$ is a $p_{X} \times p_{X}$ diagonal matrix of uniquenesses for the $\mat{X}$ variables. The $\vect{Y}$ and $\vect{Z}$ parameters are partitioned in a similar fashion. Assuming that $\mat{\Lambda}_{Y}$ and $\mat{\Lambda}_{Z}$ can be recovered from the observed blocks of the covariance matrix, it is possible to learn $\text{cov}(\vect{Y}, \vect{Z})$ as $\mat{\mat{\Sigma}}_{YZ}=\mat{\Lambda}_{Y}\mat{\Lambda}_{Z}^{\T}$. 

The file-matching problem can be approached as a low-rank matrix completion task. One may attempt to complete the partial estimate of the covariance matrix or to complete the partially observed data matrix. Generic results for matrix recovery with random missingness patterns do not give clear guidance on the feasibility of statistical file-matching with the fixed and systematic missing-data pattern in Table \ref{tab:missing} \citep{candes_2010_matrix}. 

We provide conditions for the identifiability of the factor analysis model given the missing-data pattern in the file-matching problem. The main result is that recovery of $\mat{\Sigma}_{YZ}$ is possible if the number of latent factors $q$ satisfies $q \le p_{X} , q   < (p_{X}+p_{Y})/2$, and $q < (p_{X} + p_{Z})/2$ (Theorem \ref{thm:identifiability}). Additionally, we give an EM algorithm \citep{dempster_maximum_1977} for  joint estimation of $\mat{\mat{\Lambda}}$ and $\mat{\mat{\Psi}}$ with closed form E- and M-steps.  We find that on several real datasets, maximum likelihood factor analysis gives better estimates of $\mat{\Sigma}_{YZ}$ than common existing approaches for low-rank matrix completion.

\section{Methods}
Here we review some existing approaches that can be used to estimate $\mat{\Sigma}_{YZ}$ in the statistical file-matching problem.
\subsection{Conditional independence assumption}
A common resolution to the identifiability issue in the file-matching problem is to assume that $\vect{Y}$ and $\vect{Z}$ are conditionally independent given $\vect{X}$. Under this assumption,
 ${\mat{\Sigma}}_{YZ} ={\mat{\Sigma}}_{YX}{\mat{\Sigma}}_{XX}^{-1}{\mat{\Sigma}}_{XZ}$. The conditional independence assumption is commonly invoked to justify statistical matching \citep{dorazio_2006_statistical, rassler_2002_statistical}. However, it is untestable and can have a large influence on downstream results \citep{barry_investigation_1988, rodgers_evaluation_1984}. 

\subsection{Identified set}
\label{subsec:identified_set}
To avoid making assumptions, the generative model can be treated as being partially identified \citep{gustafson_2015_bayesian}. There is a growing literature focused on developing uncertainty bounds for the matching problem that reflect the limited information in the sample \citep{dorazio_2006_categorical, conti_2012_uncertainty, conti_2016_statistical}.  For the multivariate normal model, the goal is to estimate the a feasible set of values for $\mat{\mat{\Sigma}}_{YZ}$, rather than to deliver a point estimate \citep{kadane_2001_some, moriarity_2001_statistical, ahfock_2016_partial}. The restriction that the covariance matrix be positive definite creates an identified set
\begin{align}
\Theta_{\text{I}} &= \left\lbrace \mat{\mat{\Sigma}}_{YZ} \in \mathbb{R}^{p_{Y}\times p_{Z}}: \begin{pmatrix}
\mat{\mat{\Sigma}}_{XX} & \mat{\mat{\Sigma}}_{XY} & \mat{\mat{\Sigma}}_{XZ} \\
\mat{\mat{\Sigma}}_{YX} & \mat{\mat{\Sigma}}_{YY} & \mat{\mat{\Sigma}}_{YZ} \\
\mat{\mat{\Sigma}}_{ZX} &  \mat{\mat{\Sigma}}_{ZY} &\mat{\mat{\Sigma}}_{ZZ}
\end{pmatrix} \text{is positive definite} \right\rbrace.
\label{eq:identified_set}
\end{align} 
A disadvantage of the partial identification approach is that the level of uncertainty on $\mat{\Sigma}$ can be very high.  Bounds on the parameters can be too wide for any practical interpretation.

\subsection{Low-rank matrix completion}
\label{subsec:low_rank}
Low-rank matrix completion may be applied to fill in the partial estimate of the covariance matrix or to impute the missing observations in the partially observed data matrix. To complete the estimate of the covariance matrix $\widehat{\mat{\Sigma}}$, one may consider the regularised least-squares objective
\begin{align*}
    f(\mat{M}, \mat{\Psi}) &= \dfrac{1}{2}\sum_{(i,j) \in \Omega}(\widehat{\mat{\Sigma}}_{ij}-\mat{M}_{ij}-\mat{\Psi}_{ij})^2 + \lambda \lVert \mat{M} \rVert_{*}, \\
    & \text{subject to }  \mat{M}^{\T}=\mat{M},
\end{align*}
where $\lVert \mat{M} \rVert_{*}$ is the nuclear norm of the matrix $\mat{M}$  \citep{candes_2010_matrix, mazumder_2010_spectral} and $\Omega$ contains the indices of the known elements of $\widehat{\mat{\Sigma}}$. The main barrier to attempting matrix completion on $\widehat{\mat{\Sigma}}$ is that the presence of the diagonal matrix ${\mat{\Psi}}$ along with the symmetry constraint on $\mat{M}$ necessitates modifications to existing algorithms for low-rank matrix completion.  The algorithm for symmetric low-rank matrix completion with block missingness developed in \citet{bishop_2014_deterministic} could be a useful starting point. 

A more natural approach in the file-matching problem is to use a low-rank matrix completion algorithm to impute the missing data.  Partitioning the complete-dataset $\mathcal{D}$ as
\begin{align*}
  \mathcal{D} &=  \begin{pmatrix}
     \vect{X}_{A} & \vect{Y}_{A} & \vect{Z}_{A} \\
     \vect{X}_{B} & \vect{Y}_{B} & \vect{Z}_{B}
    \end{pmatrix}
\end{align*}
where the first row corresponds to dataset A, and the second row corresponds to dataset B, the missing $\vect{Z}_{A}$ and $\vect{Y}_{B}$ observations can be estimated by applying matrix completion to $\mathcal{D}$. The modelling assumption is that complete-dataset can be expressed as $\mathcal{D}=\mathcal{D}_{0} + \mathcal{E}$ for some rank $q$ matrix $\mathcal{D}_{0}$ plus noise $\mathcal{E}$. To estimate $\mathcal{D}_{0}$, the following least squares objective can be used
\begin{align*}
    f(\mat{G}, \mat{H}) &= \dfrac{1}{2}\sum_{(i,j) \in \Omega} (\mathcal{D}_{ij}-[\mat{G}\mat{H}]_{ij})^2,
\end{align*}
where $\mat{G}$ is a $n \times q$ matrix, $\mat{H}$ is a $q \times p$ matrix and $\Omega$ contains the indices of the observed elements in $\mathcal{D}$. With $(\widehat{\mat{G}}, \widehat{\mat{H}}) \in \argmin_{(\mat{G}, \mat{H})} f(\mat{G}, \mat{H})$, the low-rank estimate of $\mathcal{D}_{0}$ is 
\begin{align}
    \widehat{\mathcal{D}}_{0} &= \widehat{\mat{G}}\widehat{\mat{H}} = \begin{pmatrix}
     \widehat{\vect{X}}_{A} & \widehat{\vect{Y}}_{A} & \widehat{\vect{Z}}_{A} \\
     \widehat{\vect{X}}_{B} & \widehat{\vect{Y}}_{B} & \widehat{\vect{Z}}_{B}
    \end{pmatrix}. \label{eq:low_rank_objective}
\end{align}
Using the low-rank reconstructions $\widehat{\vect{Z}}_{A}$ and $\widehat{\vect{Y}}_{B}$ as imputed values for the missing-data $\mat{Z}_{A}$ and $\mat{Y}_{B}$, an estimate of $\mat{\Sigma}_{YZ}$ is then given by
\begin{align*}
   \widehat{\mat{\Sigma}}_{YZ} &=  \text{cov}\left( \begin{bmatrix}
     {\vect{Y}}_{A} \\
      \widehat{\vect{Y}}_{B}
    \end{bmatrix}, \begin{bmatrix}
    \widehat{\vect{Z}}_{A} \\
    {\vect{Z}}_{B}
    \end{bmatrix} \right).
\end{align*}
The theoretical work in \cite{koltchinskii_2011} can be used to give finite-sample bounds on the reconstruction error $\lVert \mathcal{D}_{0}-\widehat{\mathcal{D}}_{0} \rVert_{2}^{2}$ under assumptions on the noise matrix $\mathcal{E}$. However, in the context of the file-matching problem, it is not immediately clear from existing theoretical work whether exact recovery of the true covariance $\mat{\Sigma}_{YZ}$ is possible in the large-sample limit.

\section{Factor models for statistical file-matching}
\subsection{Generative model}
\label{subsec:generative}
A probabilistic factor analysis model with $q$ factors has the latent variable representation
\begin{align}
\begin{pmatrix}
\vect{X} \\
\vect{Y} \\
\vect{Z}
\end{pmatrix} &=  \begin{pmatrix}
\mat{\Lambda}_{X} \\
\mat{\Lambda}_{Y} \\
\mat{\Lambda}_{Z}
\end{pmatrix} \vect{F} + \vect{\epsilon}, \label{eq:factor_model}
\end{align}
where $\vect{F} \sim \mathcal{N}(\vect{0}, \mat{I}_{q})$,  $\vect{\epsilon} \sim \mathcal{N}(\vect{0}, \mat{\Psi})$, and
\begin{align*}
\mat{\Psi}  = \begin{pmatrix}
\mat{\Psi}_{X} & \mat{0} &  \mat{0}\\
\mat{0} & \mat{\Psi}_{Y} &  \mat{0}\\
\mat{0} &   \mat{0} & \mat{\Psi}_{Z}\\
\end{pmatrix}.
\end{align*}
\cite{kamakura_2000_factor} discuss the application of factor analysis models to the statistical file-matching problem, but do not consider the identifiability of the model, or develop an algorithm for parameter estimation. We address both of these important points. We prove the identifiability of the factor analysis model in the file-matching problem under the following assumptions.

\begin{assumption}
\label{assump:low}
The matrix of factors for the $\vect{X}$ variables, $\mat{\Lambda}_{X}$, is of rank $q$.
\end{assumption}

\begin{assumption}
\label{assump:factor}
Define
\begin{align*}
\mat{\Lambda}_{A} &= \begin{pmatrix}
    \mat{\Lambda}_{X} \\
    \mat{\Lambda}_{Y}
    \end{pmatrix}, \quad    \mat{\Lambda}_{B} = \begin{pmatrix}
    \mat{\Lambda}_{X} \\
    \mat{\Lambda}_{Z}
    \end{pmatrix}.
\end{align*}
For both $\mat{\Lambda}_{A}$ and $\mat{\Lambda}_{B}$, if any row is removed, there remain two disjoint submatrices of rank $q$. 
\end{assumption}
Assumption \ref{assump:factor} is based on the sufficient conditions for the identifiability of the factor analysis model given in \cite{anderson_1956_statistical}. For the file-matching-problem, Assumption \ref{assump:factor} enforces the requirement that $q < (p_{X}+p_{Y})/2$ and $q < (p_{X}+p_{Z})/2$. Assumption 1 requires that $q \le p_{X}$.

\subsection{Identifiability}
The key to recovery of $\mat{\Sigma}_{YZ}$ is that the matrix $\mat{\Lambda}\mat{\Lambda}^{\T}$ is of rank $q \le p_{X}$. Lemma \ref{lem:low_rank_completion} shows that it is possible to complete the matrix $\mat{\Lambda}\mat{\Lambda}^{\T}$ given the missing-data pattern in Table 1. 
\begin{lemma}
\label{lem:low_rank_completion}
Suppose that Assumption \ref{assump:low} is satisfied. If all blocks other than $\mat{\Lambda}_{Y}\mat{\Lambda}_{Z}^{\T}$ and $\mat{\Lambda}_{Z}\mat{\Lambda}_{Y}^{\T}$ of 
\begin{align*}
\mat{M} &=
    \begin{pmatrix}
    \mat{\Lambda}_{X}\mat{\Lambda}_{X}^{\T} & \mat{\Lambda}_{X}\mat{\Lambda}_{Y}^{\T} & \mat{\Lambda}_{X}\mat{\Lambda}_{Z}^{\T}\\
    \mat{\Lambda}_{Y}\mat{\Lambda}_{X}^{\T} & \mat{\Lambda}_{Y}\mat{\Lambda}_{Y}^{\T} & \mat{\Lambda}_{Y}\mat{\Lambda}_{Z}^{\T} \\
    \mat{\Lambda}_{Z}\mat{\Lambda}_{X}^{\T} & \mat{\Lambda}_{Z}\mat{\Lambda}_{Y}^{\T} & \mat{\Lambda}_{Z}\mat{\Lambda}_{Z}^{\T}
    \end{pmatrix}
\end{align*}
are observed, then it is possible to recover $\mat{\Lambda}_{Y}\mat{\Lambda}_{Z}^{\T}$ from the observed elements. Consider the eigenvalue decomposition of the two sub-blocks of $\mat{M}$ that are relevant to dataset A and dataset B: 
\begin{align*}
    \mat{V}_{A}\mat{D}_{A}\mat{V}_{A}^{\T} &= \begin{pmatrix}
    \mat{\Lambda}_{X}\mat{\Lambda}_{X}^{\T} & \mat{\Lambda}_{X}\mat{\Lambda}_{Y}^{\T} \\
    \mat{\Lambda}_{Y}\mat{\Lambda}_{X}^{\T} & \mat{\Lambda}_{Y}\mat{\Lambda}_{Y}^{\T}
    \end{pmatrix}, \quad    \mat{V}_{B}\mat{D}_{B}\mat{V}_{B}^{\T} = \begin{pmatrix}
    \mat{\Lambda}_{X}\mat{\Lambda}_{X}^{\T} & \mat{\Lambda}_{X}\mat{\Lambda}_{Z}^{\T} \\
    \mat{\Lambda}_{Z}\mat{\Lambda}_{X}^{\T} & \mat{\Lambda}_{Z}\mat{\Lambda}_{Z}^{\T}
    \end{pmatrix}.
\end{align*}
The factors for the marginal models for dataset A and dataset B in the canonical rotation are
\begin{align*}
    \begin{pmatrix}\mat{\Lambda}_{X}^{A} \\ \mat{\Lambda}_{Y}^{A} \end{pmatrix}&= \mat{V}_{A}\mat{D}_{A}^{1/2}, \quad 
   \begin{pmatrix} \mat{\Lambda}_{X}^{B} \\ \mat{\Lambda}_{Z}^{B}\end{pmatrix} = \mat{V}_{B}\mat{D}_{B}^{1/2}.
 \end{align*}
To recover $\mat{\Lambda}_{Y}\mat{\Lambda}_{Z}^{\T}$ from the observed elements set
\begin{align*}
    \mat{\Lambda}_{Y}\mat{\Lambda}_{Z}^{\T} &= \mat{\Lambda}_{Y}^{A}(\mat{\Lambda}_{Z}^{B}\mat{R})^{\T},
\end{align*}
where $\mat{R}=\mat{W}\mat{Q}^{\T}$, and $\mat{W}$ and $\mat{Q}$ are the left and right singular vectors of the matrix
     $\mat{M} = (\mat{\Lambda}_{X}^{B})^{\T}{\mat{\Lambda}_{X}^{A}} = \mat{W}\mat{D}\mat{Q}^{\T}$.
\end{lemma}
The proof is given in the Appendix. \citet{bishop_2014_deterministic} establish more general results for the low-rank completion of symmetric matrices with structured missingness. Although Lemma \ref{lem:low_rank_completion} is sufficient for low-rank matrix completion problems, identifiability of the factor analysis model is a more complex issue.  Specifically, for two parameter sets $\lbrace \mat{\Lambda}, \mat{\Psi} \rbrace$ and $\lbrace \mat{\Lambda}^{*}, \mat{\Psi}^{*} \rbrace$, the equality $\mat{\Lambda}\mat{\Lambda}^{\T} + \mat{\Psi}= \mat{\Lambda}^{*}\mat{\Lambda}^{*\T} + \mat{\Psi}^{*}$  does not imply that $ \mat{\Lambda}\mat{\Lambda}^{\T}=\mat{\Lambda}^{*}\mat{\Lambda}^{*\T}$ \citep{anderson_1956_statistical, shapiro_1985_identifiability}. Assumption 2 allows us to make a statement about the identifiability of the factor analysis model in the file-matching problem.  
\begin{theorem}
\label{thm:identifiability}
Suppose that Assumptions \ref{assump:low} and \ref{assump:factor} are satisfied. Consider two parameter sets
\begin{align*}
  \vect{\Theta}_{1}=  \left\lbrace \begin{pmatrix}
    \mat{\Lambda}_{X} \\
    \mat{\Lambda}_{Y} \\
    \mat{\Lambda}_{Z}
    \end{pmatrix}, \begin{pmatrix}
    \mat{\Psi}_{X} & 0 & 0 \\
    0 & \mat{\Psi}_{Y} & 0 \\
    0 & 0 & \mat{\Psi}_{Z}
    \end{pmatrix} \right\rbrace, \quad  \vect{\Theta}_{2}=   \left\lbrace \begin{pmatrix}
    \mat{\Lambda}_{X}^{*} \\
    \mat{\Lambda}_{Y}^{*} \\
    \mat{\Lambda}_{Z}^{*}
    \end{pmatrix}, \begin{pmatrix}
    \mat{\Psi}_{X}^{*} & 0 & 0 \\
    0 & \mat{\Psi}_{Y}^{*} & 0 \\
    0 & 0 & \mat{\Psi}_{Z}^{*}
    \end{pmatrix} \right\rbrace.
\end{align*}
Then if
\begin{align}
  \begin{pmatrix}
     \mat{\Lambda}_{X}\mat{\Lambda}_{X}^{\T} + \mat{\Psi}_{X} & \mat{\Lambda}_{X}\mat{\Lambda}_{Y}^{\T} \\
      \mat{\Lambda}_{Y}\mat{\Lambda}_{X}^{\T} & \mat{\Lambda}_{Y}\mat{\Lambda}_{Y}^{\T} + \mat{\Psi}_{Y}
    \end{pmatrix} &= \begin{pmatrix}
     \mat{\Lambda}_{X}^{*}\mat{\Lambda}_{X}^{*\T} + \mat{\Psi}_{X}^{*} & \mat{\Lambda}_{X}^{*}\mat{\Lambda}_{Y}^{*\T} \\
      \mat{\Lambda}_{Y}^{*}\mat{\Lambda}_{X}^{*\T} & \mat{\Lambda}_{Y}^{*}\mat{\Lambda}_{Y}^{*\T} + \mat{\Psi}_{Y}^{*}
    \end{pmatrix}, \label{eq:lk_dataset_A} \\
    \begin{pmatrix}
     \mat{\Lambda}_{X}\mat{\Lambda}_{Z}^{\T} + \mat{\Psi}_{X} & \mat{\Lambda}_{X}\mat{\Lambda}_{Z}^{\T} \\
      \mat{\Lambda}_{Z}\mat{\Lambda}_{X}^{\T} & \mat{\Lambda}_{Z}\mat{\Lambda}_{Z}^{\T} + \mat{\Psi}_{Z}
    \end{pmatrix} &=    \begin{pmatrix}
     \mat{\Lambda}_{X}^{*}\mat{\Lambda}_{Z}^{*\T} + \mat{\Psi}_{X}^{*} & \mat{\Lambda}_{X}^{*}\mat{\Lambda}_{Z}^{*\T} \\
      \mat{\Lambda}_{Z}^{*}\mat{\Lambda}_{X}^{*\T} & \mat{\Lambda}_{Z}^{*}\mat{\Lambda}_{Z}^{*\T} + \mat{\Psi}_{Z}^{*}
    \end{pmatrix}. \label{eq:lk_dataset_B}
\end{align}
We have that the factor loadings in $\vect{\Theta}_{1}$ and $\vect{\Theta}_{2}$ are equal up to post-multiplication on the right by an orthogonal matrix. That is,
\begin{align*}
   \begin{pmatrix}
    \mat{\Lambda}_{X}\mat{\Lambda}_{X}^{\T} & \mat{\Lambda}_{X}\mat{\Lambda}_{Y}^{\T} & \mat{\Lambda}_{X}\mat{\Lambda}_{Z}^{\T} \\
    \mat{\Lambda}_{Y}\mat{\Lambda}_{X}^{\T} & \mat{\Lambda}_{Y}\mat{\Lambda}_{Y}^{\T} & \mat{\Lambda}_{Y}\mat{\Lambda}_{Z}^{\T} \\
    \mat{\Lambda}_{Z}\mat{\Lambda}_{X}^{\T} & \mat{\Lambda}_{Z}\mat{\Lambda}_{Y}^{\T} & \mat{\Lambda}_{Z}\mat{\Lambda}_{Z}^{\T}
    \end{pmatrix} &= 
   \begin{pmatrix}
    \mat{\Lambda}_{X}^{*}\mat{\Lambda}_{X}^{*\T} & \mat{\Lambda}_{X}^{*}\mat{\Lambda}_{Y}^{*\T} & \mat{\Lambda}_{X}^{*}\mat{\Lambda}_{Z}^{*\T} \\
    \mat{\Lambda}_{Y}^{*}\mat{\Lambda}_{X}^{*\T} & \mat{\Lambda}_{Y}^{*}\mat{\Lambda}_{Y}^{*\T} & \mat{\Lambda}_{Y}^{*}\mat{\Lambda}_{Z}^{*\T} \\
    \mat{\Lambda}_{Z}^{*}\mat{\Lambda}_{X}^{*\T} & \mat{\Lambda}_{Z}^{*}\mat{\Lambda}_{Y}^{*\T} & \mat{\Lambda}_{Z}^{*}\mat{\Lambda}_{Z}^{*\T}
    \end{pmatrix}.
\end{align*}
\end{theorem}
The proof is given in the Appendix. In terms of the generative model in Section \ref{subsec:generative},  
Theorem \ref{thm:identifiability} states that if parameter sets $\vect{\Theta}_{1}$ and $\vect{\Theta}_{2}$ satisfy $f(\vect{x}, \vect{y}; \vect{\Theta}_{1})= f(\vect{x}, \vect{y}; \vect{\Theta}_{2})$ and $f(\vect{x}, \vect{z}; \vect{\Theta}_{1}) =f(\vect{x}, \vect{z}; \vect{\Theta}_{2})$ then the factor loadings in $\vect{\Theta}_{1}$ and $\vect{\Theta}_{2}$ are equal up to post-multiplication on the right by an orthogonal matrix. The conclusion is that recovery of $\mat{\Sigma}_{YZ}$ in the file-matching scenario is possible given the observable sub-blocks of the population covariance matrix $\mat{\Sigma}$. Assuming that the sample covariance matrices converge in probability to the population quantities as $n_{A}, n_{B} \to \infty$, consistent estimation of $\mat{\Sigma}_{YZ}$ is possible. 

Under mild regularity conditions on the factor loadings, $q < p/2$ is sufficient for identifiability of the factor analysis model with complete-cases \citep{anderson_1956_statistical, shapiro_1985_identifiability}. In the file-matching scenario, Assumption \ref{assump:factor} enforces the stronger condition that $q < (p_{X}+p_{Y})/2$ and $q < (p_{X}+p_{Z})/2$. If the number of factors $q$ falls within the intervals $  (p_{X}+p_{Y})/2 < q < p/2$ and  $  (p_{X}+p_{Z})/2 < q < p/2$, Assumption 2 is violated. However, one may hope that the model is still identifiable, given that consistent estimation is possible with complete-cases.  

Some insight can be gained by comparing the number of equations and the number of unknowns \citep{ledermann_1937_rank}. There are $pq$ unknown parameters in $\mat{\Lambda}$ and $p$ unknown parameters in $\mat{\Psi}$. Given the symmetry in $\mat{\Sigma} = \mat{\Lambda}\mat{\Lambda}^{\T} + \mat{\Psi}$, there are $p(p+1)/2$ equations, and the adoption of a rotation constraint on $\mat{\Lambda}$ yields an additional $q(q-1)/2$ equations \citep{anderson_1956_statistical}. The number of equations minus the number of unknowns is  $C=[(p-q)^2-p-q]/2$. A useful rule of thumb is to expect the the factor analysis model to be identifiable if $C \ge 0$, and nonidentifiable if $C < 0$ \citep{ledermann_1937_rank, anderson_1956_statistical}. \cite{bekker_1997_generic} show that $C > 0$ is sufficient for the identifiability of the factor model in most circumstances. In the file-matching problem, the equality constraints due to $\mat{\Sigma}_{YZ}$ are not present and the degrees of freedom is now reduced to $C_{M}=[(p-q)^2-p-q]/2-p_{Y}p_{Z}$. If $C_{M}$ is negative and $C$ is positive it may signal a situation where the missing-data has caused the factor model to be nonidentifiable. 

Table \ref{tab:identifiablity} reports the maximum number of allowable factors $q$ using the degrees of freedom criteria $C$ and $C_{M}$, and the restrictions of Assumption \ref{assump:factor} in a file-matching problem with $p_{X}=p_{Y}=p_{Z}=p/3$. Assumption \ref{assump:factor} places stronger requirements on the number of factors $q$ than the degrees of freedom requirement $C_{M}$. The number of allowable factors using $C_{M}$ is smaller than $C$, demonstrating the information loss due to the missing-data.

\begin{table}
\caption{Maximum number of factors $q$ according to different identifiability criteria with $p_{X}=p_{Y}=p_{Z}=p/3$. The complete-cases degrees of freedom is given by $C=[(p-q)^2-p-q]/2$. The file-matching degrees of freedom is given by $C_{M}=C-p_{Y}p_{Z}$. }
\label{tab:identifiablity}
\begin{center}
\begin{tabular}{lcccccccc}
\toprule
  & & \multicolumn{7}{c}{$p$}  \\  \cline{3-9} 
Criterion & Quantity & 3 & 6 & 9 & 12 & 15 & 18 & 21 \\ 
  \hline
Degrees of freedom, complete-cases & $C$ & 1 & 3 & 5 & 7 & 10 & 12 & 15 \\ 
Degrees of freedom, file-matching & $C_{M}$ & 0 & 2 & 3 & 5 & 6 & 8 & 9 \\ 
Assumption 2& $p/3$ & 0 & 1 & 2 & 3 & 4 & 5 & 6 \\ 
\hline
\end{tabular}
\end{center}
\end{table}

\subsection{Estimation}
As mentioned in Section \ref{subsec:low_rank},  procedures for low-rank matrix completion focus on the estimation of $\mat{\Lambda}\mat{\Lambda}^{\T}$ rather than the joint estimation of $\mat{\Lambda}$ and $\mat{\Psi}$. A maximum likelihood approach to the factor analysis model facilitates joint estimation of  $\mat{\Lambda}$ and $\mat{\Psi}$. With complete-cases, the EM updates for factor analysis can be written in closed form \citep{rubin_1982_algorithms}. Let $\mat{S}$ denote the sample scatter matrix
\begin{align}
    \mat{S} &= \sum_{i=1}^{n}\begin{pmatrix}
    \vect{x}_{i} \\
    \vect{y}_{i} \\
    \vect{z}_{i}
    \end{pmatrix}
    \begin{pmatrix}
    \vect{x}_{i} \\
    \vect{y}_{i} \\
    \vect{z}_{i}
    \end{pmatrix}^{\T}. \label{eq:complete_scatter_matrix}
\end{align}
Given current parameter estimates of $\vect{\mat{\Lambda}}$ and $\mat{\mat{\Psi}}$, define $\mat{\beta} \equiv \mat{\Lambda}^{\T}(\mat{\Lambda}\mat{\Lambda}^{\T} + \mat{\Psi})^{-1}$. The M-step is
\begin{align}
    \mat{\Lambda}^{\text{new}} &= (\mat{S}\mat{\beta}^{\T})\left(  nI - n\mat{\beta}\mat{\Lambda} + \vect{\beta}\mat{S}\mat{\beta}^{\T}\right)^{-1}, \label{eq:lambda_mstep_complete} \\
    \mat{\Psi}^{\rm{new}} &= \dfrac{1}{n}\text{diag}\left(\mat{S} - \mat{\Lambda}^{\text{new}}\vect{\beta}\mat{S} \right). \label{eq:psi_mstep_complete}
\end{align}
For the file-matching problem we need to compute the conditional expectation of \eqref{eq:complete_scatter_matrix} given the missing-data pattern. The observed scatter matrices in dataset A and dataset B are 
\begin{align}
    \mat{P} &= \sum_{i=1}^{n_{A}}\begin{pmatrix}
    \vect{x}_{i} \\
    \vect{y}_{i}
    \end{pmatrix} \begin{pmatrix}
    \vect{x}_{i} \\
    \vect{y}_{i}
    \end{pmatrix}^{\T} = \begin{pmatrix}
        \mat{P}_{XX} & \mat{P}_{XY} \\
        \mat{P}_{YX} & \mat{P}_{YY}
    \end{pmatrix}, \label{eq:scatter_A} \\
 \quad \mat{T} &= \sum_{i=1}^{n_{B}}\begin{pmatrix}
    \vect{x}_{i} \\
    \vect{z}_{i}
    \end{pmatrix} \begin{pmatrix}
    \vect{x}_{i} \\
    \vect{z}_{i}
    \end{pmatrix}^{\T} = \begin{pmatrix}
        \mat{T}_{XX} & \mat{T}_{XZ} \\
        \mat{T}_{ZX} & \mat{T}_{ZZ}
    \end{pmatrix}. \label{eq:scatter_B}
\end{align}
To determine the conditional expectation of the complete-data sufficient statistic, for current parameter estimates of $\mat{\Lambda}$ and $\mat{\Psi}$ define
\begin{align*}
    \vect{\omega} &= \begin{pmatrix}
        \vect{\omega}_{X} & \vect{\omega}_{Y}
    \end{pmatrix}= \begin{pmatrix}
        \mat{\Lambda}_{Z}\mat{\Lambda}_{X}^{\T}& \mat{\Lambda}_{Z}\mat{\Lambda}_{Y}^{\T}
    \end{pmatrix} \begin{pmatrix}
        \mat{\Lambda}_{X}\mat{\Lambda}_{X}^{\T} + \mat{\Psi}_{X} & \mat{\Lambda}_{X}\mat{\Lambda}_{Y}^{\T} \\
        \mat{\Lambda}_{Y}\mat{\Lambda}_{X}^{\T} & \mat{\Lambda}_{Y}\mat{\Lambda}_{Y}^{\T} + \mat{\Psi}_{Y}
    \end{pmatrix}^{-1},  \\
    \vect{\alpha} &= \begin{pmatrix}
        \vect{\alpha}_{X} & \vect{\alpha}_{Z}
    \end{pmatrix}=\begin{pmatrix}
        \mat{\Lambda}_{Y}\mat{\Lambda}_{X}^{\T}& \mat{\Lambda}_{Y}\mat{\Lambda}_{Z}^{\T}
    \end{pmatrix} \begin{pmatrix}
        \mat{\Lambda}_{X}\mat{\Lambda}_{X}^{\T} + \mat{\Psi}_{X} & \mat{\Lambda}_{X}\mat{\Lambda}_{Z}^{\T} \\
        \mat{\Lambda}_{Z}\mat{\Lambda}_{X}^{\T} & \mat{\Lambda}_{Z}\mat{\Lambda}_{Z}^{\T} + \mat{\Psi}_{Z}
    \end{pmatrix}^{-1},
\end{align*}
and the conditional covariance matrices \begin{align*}
\mat{\Sigma}_{Z \mid XY} &= \begin{pmatrix}
    \mat{\Lambda}_{Z}\mat{\Lambda}_{Z}^{\T} + \mat{\Psi}_{Z}
\end{pmatrix} - \begin{pmatrix}
    \mat{\Lambda}_{Z}\mat{\Lambda}_{X}^{\T} & \mat{\Lambda}_{Z}\mat{\Lambda}_{Y}^{\T}
\end{pmatrix}\begin{pmatrix}
    \mat{\Lambda}_{X}\mat{\Lambda}_{X}^{\T} + \mat{\Psi}_{X} & \mat{\Lambda}_{X}\mat{\Lambda}_{Y}^{\T} \\
    \mat{\Lambda}_{Y}\mat{\Lambda}_{X}^{\T} & \mat{\Lambda}_{Y}\mat{\Lambda}_{Y}^{\T} + \mat{\Psi}_{Y}
\end{pmatrix}^{-1}   \begin{pmatrix}
    \mat{\Lambda}_{X}\mat{\Lambda}_{Z}^{\T} \\
    \mat{\Lambda}_{Y}\mat{\Lambda}_{Z}^{\T}
\end{pmatrix}, \\
\mat{\Sigma}_{Y \mid XZ} &= \begin{pmatrix}
    \mat{\Lambda}_{Y}\mat{\Lambda}_{Y}^{\T} + \mat{\Psi}_{Y}
\end{pmatrix} - \begin{pmatrix}
    \mat{\Lambda}_{Y}\mat{\Lambda}_{X}^{\T} & \mat{\Lambda}_{Y}\mat{\Lambda}_{Z}^{\T}
\end{pmatrix}\begin{pmatrix}
    \mat{\Lambda}_{X}\mat{\Lambda}_{X}^{\T} + \mat{\Psi}_{X} & \mat{\Lambda}_{X}\mat{\Lambda}_{Z}^{\T} \\
    \mat{\Lambda}_{Z}\mat{\Lambda}_{X}^{\T} & \mat{\Lambda}_{Z}\mat{\Lambda}_{Z}^{\T} + \mat{\Psi}_{Z}
\end{pmatrix}^{-1}   \begin{pmatrix}
    \mat{\Lambda}_{X}\mat{\Lambda}_{Y}^{\T} \\
    \mat{\Lambda}_{Z}\mat{\Lambda}_{Y}^{\T}
\end{pmatrix}.
\end{align*}
Using the augmented scatter matrices
\begin{align*}
    \widetilde{\mat{P}}&= \begin{pmatrix}
\mat{P}_{XX} & \mat{P}_{XY} & \mat{P}_{XX}\vect{\omega}_{X}^{\T} + \mat{P}_{XY}\vect{\omega}_{Y}^{\T} \\
\mat{P}_{YX}& \mat{P}_{YY} & \mat{P}_{YX}\vect{\omega}_{X}^{\T} + \mat{P}_{YY}\vect{\omega}_{Y}^{\T} \\
\vect{\omega}_{Y}\mat{P}_{YX} + \vect{\omega}_{X}\mat{P}_{XX} & \vect{\omega}_{Y}\mat{P}_{YY} + \vect{\omega}_{X}\mat{P}_{XY} & \vect{\omega}\mat{P}\vect{\omega}^{\T} + n_{A}\mat{\Sigma}_{Z \mid XY}
    \end{pmatrix},\\
       \widetilde{\mat{T}}&= \begin{pmatrix}
\mat{T}_{XX} & \mat{T}_{XX}\vect{\alpha}_{X}^{\T} + \mat{T}_{XZ}\vect{\alpha}_{Z}^{\T}& \mat{T}_{XZ} \\
\vect{\alpha}_{Z}\mat{T}_{ZX} + \vect{\alpha}_{X}\mat{T}_{XX} &\vect{\alpha}\mat{T}\vect{\alpha}^{\T}  + n_{B} \mat{\Sigma}_{Y \mid XZ}& \vect{\alpha}_{Z}\mat{T}_{ZZ} + \vect{\alpha}_{X}\mat{T}_{XZ}  \\
\mat{T}_{ZX}  &\mat{T}_{ZX}\vect{\alpha}_{X}^{\T} + \mat{T}_{ZZ}\vect{\alpha}_{Z}^{\T}  & \mat{T}_{ZZ}
    \end{pmatrix},
\end{align*}
the expected value of the complete-data sufficient statistic is given by
\begin{align*}
\widetilde{\mat{S}} &=\widetilde{\mat{P}} +  \widetilde{\mat{T}}.
\end{align*}
Given current parameter estimates of ${\mat{\Lambda}}, {\mat{\Psi}}$, define $\mat{\beta} \equiv \mat{\Lambda}^{\T}(\mat{\Lambda}\mat{\Lambda}^{\T} + \mat{\Psi})^{-1}$. The M-step is then
\begin{align}
    \mat{\Lambda}^{\text{new}} &= (\widetilde{\mat{S}}\mat{\beta}^{\T})\left(  n\mat{I}-n\vect{\beta}\mat{\Lambda} + \vect{\beta}\widetilde{\mat{S}}\vect{\beta}^{\T}\right)^{-1}, \\
    \mat{\Psi}^{\rm{new}} &= \dfrac{1}{n}\text{diag}\left(\widetilde{\mat{S}} - \mat{\Lambda}^{\text{new}}\vect{\beta}\tilde{\mat{S}} \right).
\end{align}
The EM algorithm can be run using the sample covariance matrices from dataset A and dataset B, and so it is not necessary to have the original observations. The multivariate normality assumption used to develop the algorithm is not crucial, as long as the population covariance matrix fits the factor analysis model $\mat{\Sigma}=\mat{\Lambda}\mat{\Lambda}^{\T} + \mat{\Psi}$ \citep{browne_1984_asymptotically}. 

\subsection{Model selection}
In practice, the number of factors $q$ must be determined. The Bayesian information criterion \citep{schwarz_1978_estimating} is a generic tool for model selection that has been well explored in factor analysis for selecting $q$ \citep{preacher_2013_choosing}. The Bayesian information criterion for a model $\mathcal{M}$ with likelihood function $L$ and parameter $\vect{\theta}$ is given by
\begin{align}
    \text{BIC}(\mathcal{M}) &= -2\log L(\vect{\widehat{\theta}}) +  d \log n, \label{eq:bic_generic}
\end{align}
where $\widehat{\vect{\theta}}$ is the maximum likelihood estimate,  $d$ is the number of free parameters in the model $\mathcal{M}$, and $n$ is the sample size \citep{schwarz_1978_estimating}. For factor analysis with complete-cases the log-likelihood is
\begin{align*}
  \log L(\mat{\Lambda}, \mat{\Psi}) &=   -\dfrac{n}{2} \log \lvert \mat{\Sigma} \rvert - \dfrac{1}{2}\text{trace}\lbrace \mat{\Sigma}^{-1}\mat{S} \rbrace - \dfrac{np}{2}\log 2\pi,
\end{align*}
where $\mat{\Sigma} = \mat{\Lambda}\mat{\Lambda}^{\T}+\mat{\Psi}$ and $\mat{S}$ is the sample scatter matrix \eqref{eq:complete_scatter_matrix}. The Bayesian information criterion for factor analysis with complete-cases is therefore
\begin{align*}
    \text{BIC}(q) &= -2 \log  L(\widehat{\mat{\Lambda}}, \widehat{\mat{\Psi}}) +(qp+p-q(q-1)/2)\log n,
\end{align*}
where the number of free parameters is obtained by subracting the number of rotation constraints $q(q-1)/2$ from the nominal number of parameters $qp+p$. 

For model selection in missing-data problems it has been suggested to use the observed-data log-likelihood in place of the complete-data log-likelihood in the standard definition of the Bayesian information criterion \eqref{eq:bic_generic} \citep{ibrahim_2008_model}. This is the strategy we propose to use for determining the number of factors $q$ in the file-matching problem. The log-likelihood functions for each dataset are given by
\begin{align*}
  \log L_{A}(\mat{\Sigma}, \mat{\Psi}) &=   -\dfrac{n_{A}}{2} \log \lvert \mat{\Sigma}_{A} \rvert - \dfrac{1}{2}\text{trace}\lbrace \mat{\Sigma}_{A}^{-1}\mat{P} \rbrace - \dfrac{n_{A}p_{A}}{2} \log 2\pi, \\
   \log L_{B}(\mat{\Sigma}, \mat{\Psi}) &=   -\dfrac{n_{B}}{2} \log \lvert \mat{\Sigma}_{B} \rvert - \dfrac{1}{2}\text{trace}\lbrace \mat{\Sigma}_{B}^{-1}\mat{T} \rbrace - \dfrac{n_{B}p_{B}}{2} \log 2\pi,
\end{align*}
where $p_{A}=p_{X}+p_{Y}$, $p_{B}=p_{X}+p_{Z}$, $\mat{P}$ is the observed scatter matrix from dataset A \eqref{eq:scatter_A}, $\mat{T}$ is the observed scatter matrix from dataset B \eqref{eq:scatter_B} and
\begin{align*}
   \mat{\Sigma}_{A}  &=   \begin{pmatrix}
     \mat{\Lambda}_{X}\mat{\Lambda}_{X}^{\T} + \mat{\Psi}_{X} & \mat{\Lambda}_{X}\mat{\Lambda}_{Y}^{\T} \\
      \mat{\Lambda}_{Y}\mat{\Lambda}_{X}^{\T} & \mat{\Lambda}_{Y}\mat{\Lambda}_{Y}^{\T} + \mat{\Psi}_{Y}
    \end{pmatrix}, \\
    \mat{\Sigma}_{B} &=    \begin{pmatrix}
     \mat{\Lambda}_{X}\mat{\Lambda}_{X}^{\T} + \mat{\Psi}_{X} & \mat{\Lambda}_{X}\mat{\Lambda}_{Z}^{\T} \\
      \mat{\Lambda}_{Z}\mat{\Lambda}_{X}^{\T} & \mat{\Lambda}_{Z}\mat{\Lambda}_{Z}^{\T} + \mat{\Psi}_{Z}
    \end{pmatrix}.
\end{align*}
The Bayesian information criterion using the observed-data log-likelihood is then
\begin{align*}
    \text{BIC}(q) &=  -2\{ \log L_{A}(\widehat{\mat{\Lambda}}, \widehat{\mat{\Psi}}) + \log L_{B}(\widehat{\mat{\Lambda}}, \widehat{\mat{\Psi}})\}  +(qp+p-q(q-1)/2)\log n.
\end{align*}
\section{Simulation}
Here we investigate the necessity of Assumption 2 for the identifiability of the factor model by focusing on the noiseless setting where the exact covariance matrix $\mat{\Sigma}$ is known with the exception of $\mat{\Sigma}_{YZ}$. We considered a file-matching scenario with $p_{X}=p_{Y}=p_{Z}=4$. As seen in Table \ref{tab:identifiablity}, Assumption \ref{assump:factor} enforces stricter requirements than the degrees of freedom criterion based on $C_{M}=[(p-q)^2-p-q]/2-p_{Y}p_{Z}$. We simulated three different covariance matrices with $p=12$, one from a $q=3$ factor model, one from a $q=4$ factor model, and one from a $q=5$ factor model. Elements of $\Lambda$ were sampled from a $\mathcal{N}(2, 1)$ distribution. The $j$th diagonal element of $\Psi$ was set as $D_{j}^{2}$, where $D_{j} \sim \mathcal{N}(3, 0.01)$. Assumption 2 holds only for $q=3$, but the degrees of freedom $C_{M}$ is non-negative for $q=3,4,5$.

\begin{figure}
\centering
\includegraphics[width=0.66\textwidth]{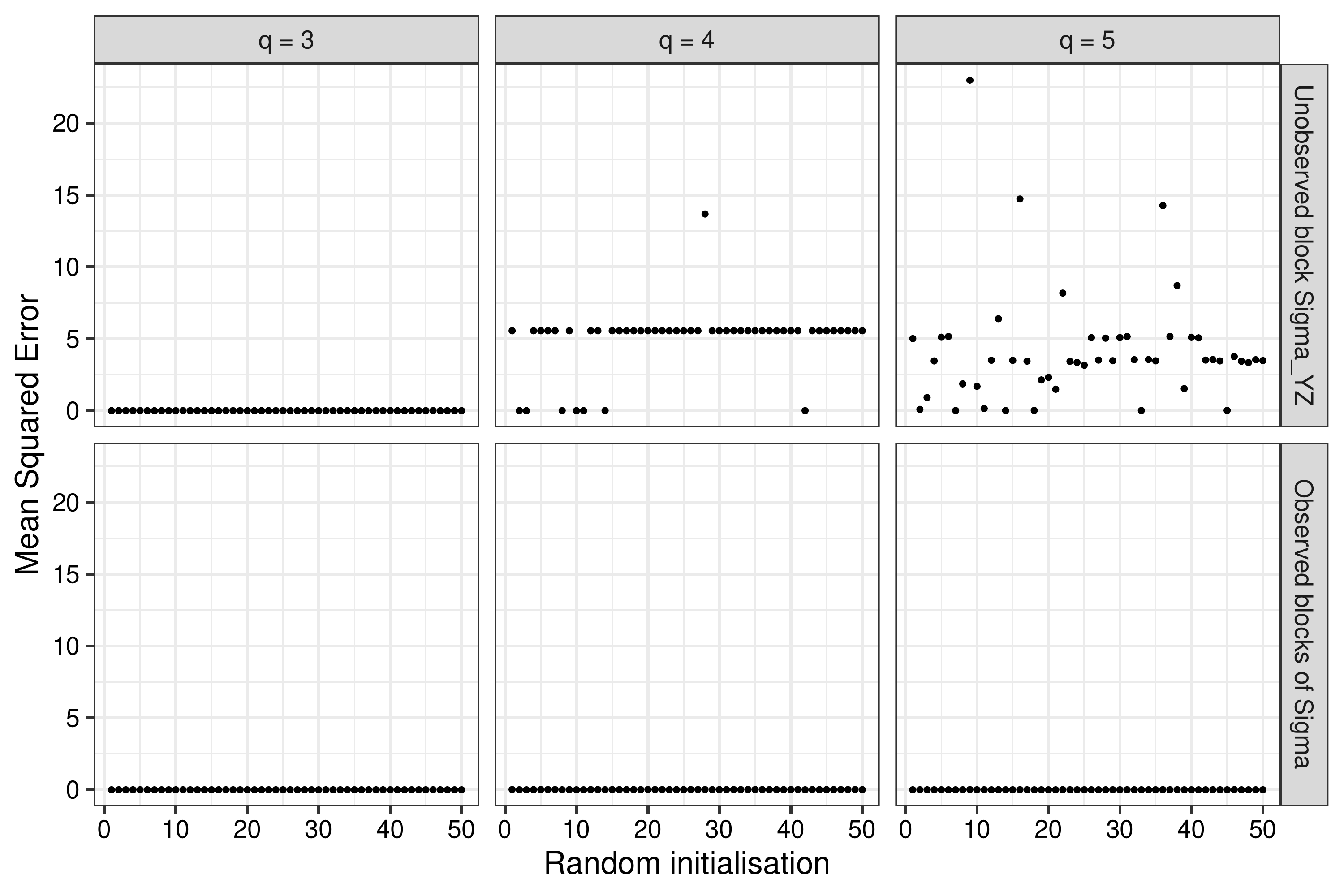}
\caption{Estimation errors in a simulated file-matching scenario with $p=12$ and $p_{X}=p_{Y}=p_{Z}=4$. The EM algorithm was run for 10000 iterations,  panels show the estimation error from different random initialisations. Bottom row: for $q=3,4$, and $5$ the reconstruction error of the observed blocks of $\Sigma$ is zero across all initialisations. Top row: for $q=4$ and $5$, the random initialisations appear to lead to different estimates of $\mat{\Sigma}_{YZ}$, suggesting multiple factor solutions. For $q=3$, Assumption \ref{assump:factor} is satisfied, and the factor model is identifiable. For $q=3$ the estimation error of $\mat{\Sigma}_{YZ}$ is zero across all initialisations. 
\label{fig:simulation}}
\end{figure}

For each covariance matrix, we fit a factor model using the EM algorithm for 10000 iterations using a random initial value $\Lambda$, where each element was sampled from a $\mathcal{N}(0,1)$ distribution. The correct number of factors was used for each of the three test matrices. Given the estimate $\widehat{\mat{\Sigma}}$, we computed the mean squared error over the elements of $\mat{\Sigma}_{YZ}$, that is $\lVert \widehat{\mat{\Sigma}}_{YZ} - \mat{\Sigma}_{YZ} \rVert^{2}/(p_{Y}p_{Z})$. We also computed the mean squared error for the observed blocks of $\mat{\Sigma}$. This procedure was repeated fifty times. Figure \ref{fig:simulation} displays the results over the fifty random initialisations for each of the three covariance matrices $(q=3,4,5)$.

The top row shows the estimation errors for $\mat{\Sigma}_{YZ}$, the bottom row shows the estimation errors for the observed blocks of $\mat{\Sigma}$. Looking at the bottom row, for $q=3,4$, and $5$ the EM algorithm is able to reconstruct the observed blocks of $\mat{\Sigma}$ over each random initialisation. From the top row, for $q=3$, the choice of initial values does not affect the final estimate of $\mat{\Sigma}_{YZ}$, and the estimation error is zero in each trial. For $q=4$ and $5$, the choice of initial values influences the final estimate of $\mat{\Sigma}_{YZ}$. As the reconstruction error for the observed blocks of $\mat{\Sigma}$ is zero, this suggests there are multiple factor solutions and the model is not identifiable. The divergent behaviour when comparing $q=3$ to $q=4$ and $5$ is notable, as Assumption \ref{assump:factor} is only satisfied for $q=3$. The degrees of freedom $C_{M}$ are non-negative for $q=4$ and $5$, but this does not lead to a well-behaved estimator. In this simulation,  Assumption \ref{assump:factor} appears to give better guidelines for the stable recovery of $\mat{\Sigma}_{YZ}$ compared to $C_{M}$.

\section{Data application}
\subsection{Estimation}
In this section we compare the performance of various algorithms for the estimation of $\mat{\Sigma}_{YZ}$. Existing algorithms for low-rank matrix completion were implemented using the R package \texttt{filling} \citep{you_filling_2020}. We also considered the standard conditional independence assumption (CIA). We tested the algorithms on the following datasets.
\begin{itemize}
    \item Sachs dataset. This dataset consists of results from 11 flow cytometry experiments measuring the same set of 12 variables. The measurements were log transformed. Data from each experiment was centered, then combined and scaled so that each variable had unit variance. 
    \item Wisconsin breast cancer dataset. The original observations in the malignant class were centered and scaled so that each variable had unit variance. 
    \item Abalone dataset. The original observations in the female class were centered and scaled so that each variable had unit variance. 
\end{itemize}
For each dataset we considered matching scenarios with $p_{X}$ common variables, $p_{Y}$ unique variables in dataset A, and $p_{Z}$ unique variables in dataset B. This defines partitions of the complete-dataset. Table  \ref{tab:examples} gives the file-matching scenarios considered for each dataset, as well as the number of factors $q$ that were used in the fitting of the factor analysis model. The final columns indicate whether the degrees of freedom criteria $C$ and $C_{M}$, and Assumption 2 are satisfied for the chosen number of factors $q$. Assumption 2 is violated on the Sachs dataset, but the degrees of freedom $C_{M}$ is positive.

\begin{table}
\caption{Data examples. $C$ represents the degrees of freedom given complete cases, $C_{M}$ represents the degrees of freedom in the file-matching scenario. The final column states if it  possible to satisfy Assumption 2 given the file-matching scenario.}
\label{tab:examples}
\begin{center}
\begin{tabular}{@{}lcccccccc@{}}
\toprule
Dataset &\multicolumn{1}{c}{$p$}& \multicolumn{1}{c}{$p_{X}$} &  \multicolumn{1}{c}{$p_{Y}$} & \multicolumn{1}{c}{$p_{Z}$}& \multicolumn{1}{c}{$q$} & \multicolumn{1}{c}{$C \ge 0$}  & \multicolumn{1}{c}{$C_{M} \ge 0$} & Assumption 2 \\ \midrule
Sachs  & 11 & 4 & 4 & 3 & 4 & Yes & Yes & No\\
Wisconsin & 10 & 3 & 3 & 4 & 2 & Yes &  Yes & Yes\\
Abalone & 8 & 1 & 4 & 3 & 1 & Yes &  Yes & Yes\\
\bottomrule
\end{tabular}
\end{center}
\end{table}

In each simulation, variables were randomly allocated to the $\mat{X}$, $\vect{Y}$, and $\vect{Z}$ groups. We then used different algorithms to estimate $\mat{\Sigma}_{YZ}$. The mean squared error over elements of $\mat{\Sigma}_{YZ}$ was then recorded for each algorithm, that is $\lVert \widehat{\mat{\Sigma}}_{YZ} - \mat{\Sigma}_{YZ} \rVert^{2}/(p_{Y}p_{Z})$. This process was repeated 100 times. The EM algorithm was run for 2000 iterations in each simulation. We considered two different initial value settings for the EM algorithm, initialisation using the factor analysis solution from the full covariance matrix, and random initialisation of $\mat{\Lambda}$ by sampling from a standard normal distribution. For the random initialisation protocol we took one hundred samples of $\mat{\Lambda}$ followed by a short run of the EM algorithm for fifty iterations. The parameters with the highest log-likelihood after the fifty iterations were then used in the longer run. In each simulation we also fit a factor model using the complete-dataset with no missing data to provide a reference point for the goodness-of-fit of the factor model.

\begin{figure}
\includegraphics[width=\textwidth]{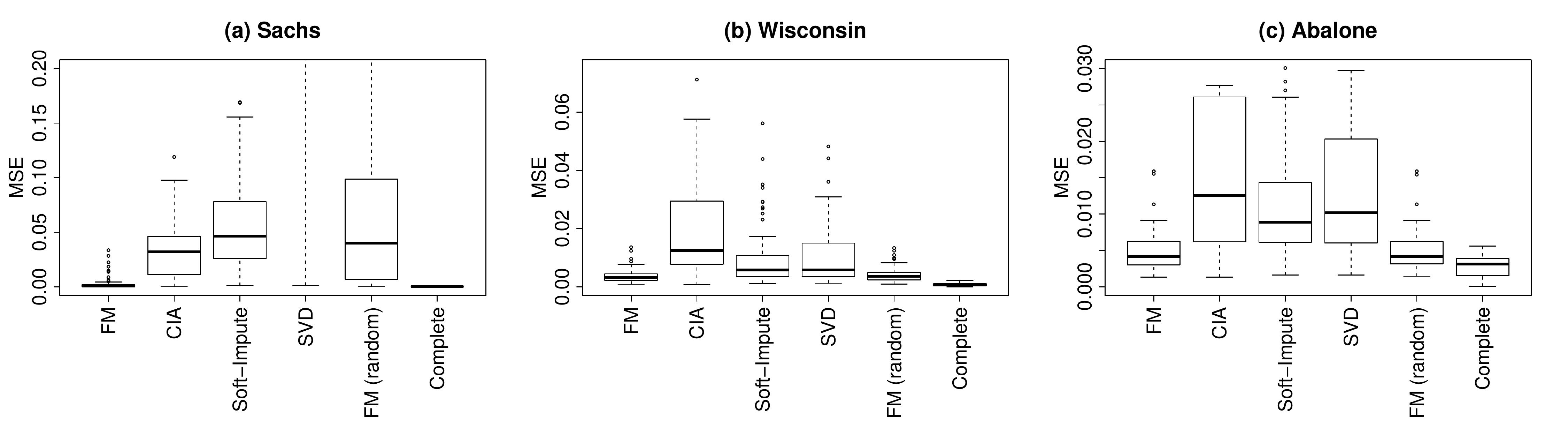}
\caption{Accuracy of estimates of $\mat{\Sigma}_{YZ}$ over 100 variable permutations for different algorithms. Complete refers to estimation of a factor analysis model using complete-cases with no missing data. 
\label{fig:fm}}
\end{figure}

Figure \ref{fig:fm}  compares the results using the different algorithms. The results for factor analysis model with the favourable initial values are shown as FM, the results for the factor analysis model with random initialisation are shown as FM (random). The factor model with good initial values (FM) has the lowest median error across each dataset (Table \ref{tab:fm}). The errors on the Sachs dataset and the Wisconsin breast cancer dataset are particularly small.  

The results for the factor model with random initialisation (FM (random)) are very similar to those of FM using the favourable initial values except on the Sachs dataset.  On the Sachs dataset, Assumption 2 is violated and there may be multiple factor solutions. In Figure \ref{fig:fm} (a), the errors for the factor model with random initialisation (FM (random)) are much more dispersed than the factor model with the good initial values (FM). Although there are sufficient degrees of freedom $(C_{M} \ge 0)$, it appears that the model may not be identifiable. With good initial values, the EM algorithm appears to converge to a local mode that gives a good estimate of $\Sigma_{YZ}$. 

The interquartile range of the error for each algorithm over the one hundred variable permutations is given in Table \ref{tab:fm}. The factor analysis model (FM) has the smallest interquartile range on each dataset, closely followed by the factor model using random initialisations with the exception of the Sachs dataset.

\begin{table}[ht]
\centering
\caption{Median error and interquartile range in estimation of $\mat{\Sigma}_{YZ}$ across 100 variable permutations. The factor analysis model (FM) has the lowest median error and interquartile range on each dataset. The factor analysis model with random initial values (FM (random)) has comparable performance in each dataset except for Sachs, where Assumption 2 is violated.}
\begin{tabular}{@{}rlllllll@{}}
  \hline
 Dataset &Value& FM & CIA & Soft-Impute & SVD & FM (random) & Complete \\ 
  \hline
Sachs & Median&  \textbf{0.0006} & 0.0320 & 0.0465 & 3.1174 & 0.0402 & 0.0001 \\ 
      & IQR &  \textbf{0.0017} & 0.0351 & 0.0518 & 27.822 & 0.0907 & 0.0001 \\ &&&&&& \\
  Wisconsin & Median& \textbf{0.0032} & 0.0125 & 0.0058 & 0.0059 & 0.0036 & 0.0006 \\ 
  & IQR &  \textbf{0.0023} & 0.0216 & 0.0073 & 0.0113 & 0.0025 & 0.0008 \\ &&&&&& \\
  Abalone & Median & \textbf{0.0042} & 0.0125 & 0.0089 & 0.0102 & \textbf{0.0042} & 0.0031 \\ 
  & IQR & \textbf{0.0031} & 0.0198 & 0.0081 & 0.0142 & \textbf{0.0031} & 0.0024 \\
   \hline
\end{tabular}
\label{tab:fm}
\end{table}

\subsection{Model selection}
We also performed some experiments to assess the behavior of the Bayesian information criterion for determining the number of factors $q$.  The following datasets and variable partitions were used in the experiments

\begin{itemize}
    \item Reise dataset. This dataset is a $16 \times 16$ correlation matrix for mental health items. The variables were partitioned as $p_{X}=6, p_{Y}=5$, and $p_{Z}=5$. 
    \item Harman dataset. This dataset is a $24 \times 24$ correlation matrix for psychological tests.  The variables were partitioned as $p_{X}=5, p_{Y}=10$, and $p_{Z}=9$. 
    \item Holzinger dataset. This dataset is a $14 \times 14$ correlation matrix for mental ability scores. The variables were partitioned as $p_{X}=6, p_{Y}=4$, and $p_{Z}=4$. 
    \item Simulated from a factor model with $q=6$ and $p=100$. Elements of $\Lambda$ were sampled from a $\mathcal{N}(2, 1)$ distribution. The $j$th diagonal element of $\Psi$ was set as $D_{j}^{2}$, where $D_{j} \sim \mathcal{N}(3, 0.01)$. The covariance matrix $\mat{\Sigma}=\mat{\Lambda}\mat{\Lambda}^{\T} + \mat{\Psi}$ was then standardised to a correlation matrix. We simulated one dataset with  $n_{A}=n_{B}=50$ observations and one dataset of $n_{A}=n_{B}=500$ observations.  The variables were partitioned as $p_{X}=6, p_{Y}=47, p_{Z}=47$.
\end{itemize}

\begin{figure}
\includegraphics[width=\textwidth]{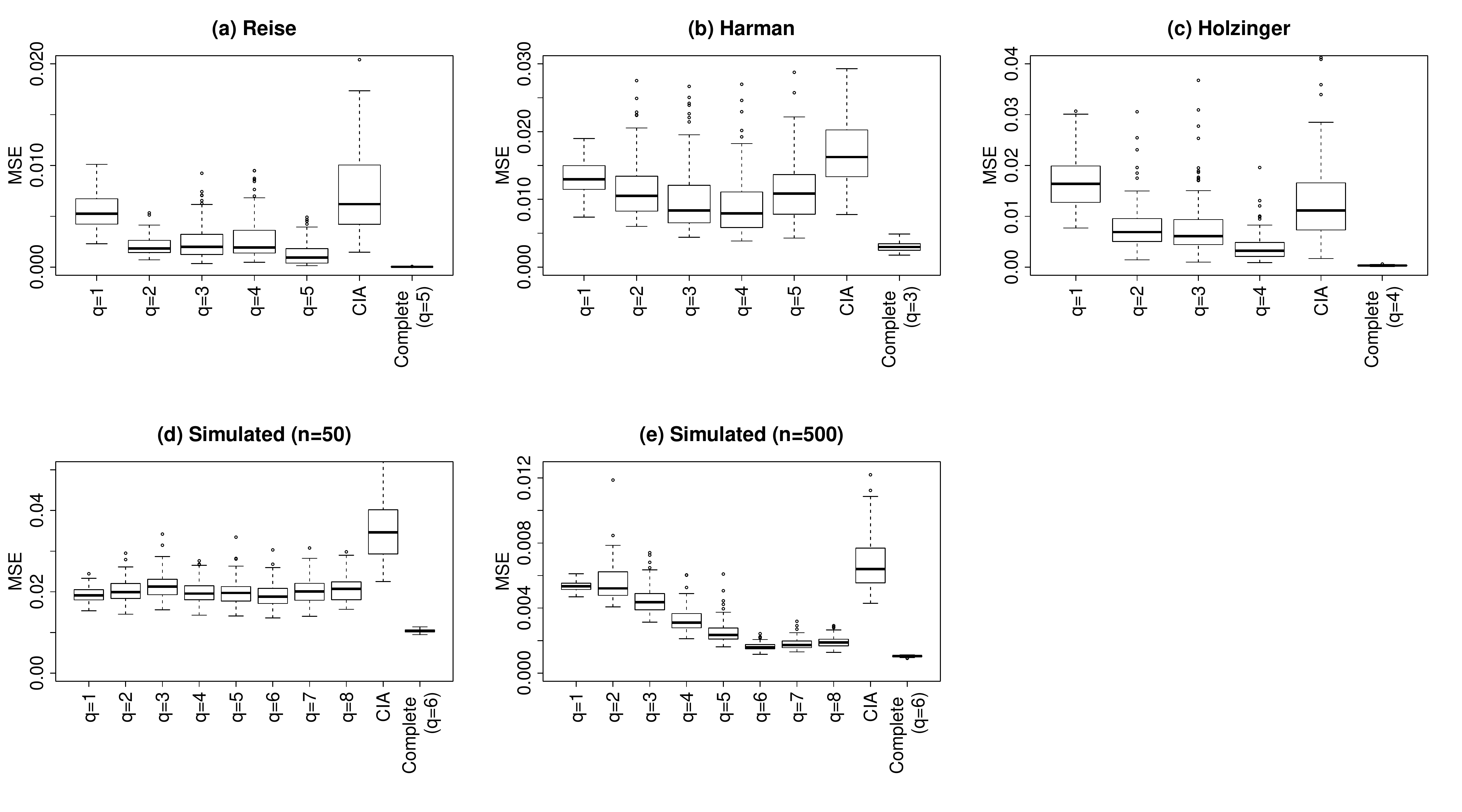}
\caption{Accuracy of estimates of $\mat{\Sigma}_{YZ}$ over 100 variable permutations for different algorithms. Complete refers to estimation of a factor analysis model using complete-cases with no missing data. 
\label{fig:model_selection}}
\end{figure}

In each simulation, variables were randomly allocated to the $\vect{X}$, $\vect{Y}$ and $\vect{Z}$ groups with $n_{A}=n_{B}=n/2$. We then computed the BIC for a range of values for $q$ and recorded the optimal value $q^{*}$.  This process was repeated 100 times. In each of the simulations we considered the range of values for $q$ such that Assumptions 1 and 2 were satisfied. Table \ref{tab:model} reports the number of times a $q$ factor model was chosen using the BIC over the 100 replications. The optimal number of factors according to the BIC using complete-cases is given as $q_{c}^{*}$ in Table \ref{tab:model}. The missing-data appears to lead to the BIC acting conservatively, selecting a number of factors less than or equal to the number chosen with complete-cases. The correct number of factors $q=6$ is selected for the simulated datasets when using complete-cases. For the simulated dataset with $n=50$, the BIC selects  fewer than six factors in each trial. For the simulated dataset with $n=500$ the BIC selects the correct number of factors $q=6$ in each replication.  Figure \ref{fig:model_selection} shows boxplots of the estimation error  $\lVert \widehat{\mat{\Sigma}}_{YZ} - \mat{\Sigma}_{YZ} \rVert^{2}/(p_{Y}p_{Z})$ for different values of $q$. The errors from using the conditional independence assumption and from fitting a factor model with complete-cases are also reported for comparison. For almost all values of $q$, the median error using the factor model is lower than that of the conditional independence model. 

\begin{table}[ht]
\centering
\caption{Simulation results for model selection using the Bayesian information criterion. The optimal number of factors using complete-cases is denoted as $q_{c}^{*}$. Entries represent the number of times a $q$ factor model was chosen using the BIC across the 100 variable permutations. A dash indicates that the $q$ factor model would violate the identification conditions in Assumptions 1 and 2.}
\begin{tabular}{lrrrrrrrrr}
  \hline
Dataset &  $q_{\text{c}}^{*}$ & $q=1$ & $q=2$ & $q=3$ & $q=4$ & $q=5$ & $q=6$ & $q=7$ & $q=8$ \\ 
  \hline
Reise &   8 &   0 &   1 &   1 &  18 &  80 &   - &   - &   - \\ 
  Harman &   3 &  62 &  37 &   1 &   0 &   0 &   - &   - &   - \\ 
  Holzinger &   4 &   0 &  60 &  37 &   3 &   - &   - &   - &   - \\ 
  Simulated $(n=50)$ &   6 &   5 &  51 &  41 &   3 &   0 &   0 &   0 &   0 \\ 
  Simulated $(n=500)$ &   6 &   0 &   0 &   0 &   0 &   0 & 100 &   0 &   0 \\ 
   \hline
\end{tabular}
\label{tab:model}
\end{table}

\section{Discussion}
\label{s:discuss}

Technological and design constraints can prevent investigators from collecting a full dataset on all variables of interest. The statistical file-matching problem is an important data-fusion task where joint observations on the full set of variables are not available. Factor analysis models are useful as they can remain identifiable despite the missing-data pattern of the file-matching problem. The factor analysis approach is a useful alternative to the conditional independence assumption, as it is less restrictive and testable. Estimation of the factor analysis model can be carried out via the EM algorithm.

Although factor analysis and low-rank matrix completion are related, the identifiability of the factor analysis model $\mat{\Sigma}=\mat{\Lambda}\mat{\Lambda}^{\T} + \mat{\Psi}$ requires additional assumptions compared to low-rank matrix completion due to the diagonal matrix $\mat{\Psi}$. As the assumption that $q < (p_{X}+p_{Y})/2$ and $q < (p_{X}+p_{Z})/2$ may be unnecessarily strong, it is of interest to establish the weakest conditions that ensure the identifiability of the factor analysis model in the file-matching problem.  In many applications of file-matching, procedures for generating uncertainty bounds when the model is partially identified supply useful information \citep{conti_2016_statistical}. Further work may explore characterisations of the identified set for the factor analysis model when the number of latent factors $q$ exceeds the maximum number for identifiability. 

It is also of interest to relax the assumption that both datasets are samples from the same homogeneous population with the joint model $f(\vect{x}, \vect{y}, \vect{z}; \vect{\theta})$. One possible avenue is to embed the file-matching problem in a hierarchical model. A dataset specific random effect can be applied to the parameters in dataset A and dataset B, so that there is shared component across datasets and a unique component within each dataset. Samples in dataset A are from the distribution $f(\vect{x}, \vect{y}; \vect{\theta}_{A})$, and samples from dataset B are from the distribution $f(\vect{x}, \vect{z}; \vect{\theta}_{B})$ where $\vect{\theta}_{A}=\vect{\theta}_{0} + \delta_{A}$ and $\vect{\theta}_{B}=\vect{\theta}_{0} + \delta_{B}$ where $\delta_{A}$ and $\delta_{B}$ are random effect terms. The added flexibility of the hierarchical model can allow for situations where datasets A and B are from related but not necessarily identical populations, a common scenario in data integration. Model identifiability with heterogeneous data sources is an interesting and challenging problem, and the results here may serve as useful groundwork. 

\section*{Acknowledgements}
This research was funded by the Australian Government through the
Australian Research Council (Project Numbers DP170100907 and IC170100035).

\bibliographystyle{rss}
\bibliography{bibliography}

\appendix

\section{Appendix}
\setcounter{equation}{0}
\renewcommand{\theequation}{A.\arabic{equation}}

\subsection{Proof of Lemma \ref{lem:low_rank_completion}}
Due to the rotational invariance of the factor model, we have that
\begin{align}
    \begin{bmatrix}
    \mat{\Lambda}_{X}^{A} \\
    \mat{\Lambda}_{Y}^{A} \\
    \mat{\Lambda}_{Z}^{A}
    \end{bmatrix} &=   \begin{bmatrix}
    \mat{\Lambda}_{X} \\
    \mat{\Lambda}_{Y} \\
    \mat{\Lambda}_{Z}
    \end{bmatrix}\mat{R}_{1}, \quad     \begin{bmatrix}
    \mat{\Lambda}_{X}^{B} \\
    \mat{\Lambda}_{Y}^{B} \\
    \mat{\Lambda}_{Z}^{B}
    \end{bmatrix} =   \begin{bmatrix}
    \mat{\Lambda}_{X} \\
    \mat{\Lambda}_{Y} \\
    \mat{\Lambda}_{Z}
    \end{bmatrix}\mat{R}_{2},
    \quad       \begin{bmatrix}
    \mat{\Lambda}_{X}^{B} \\
    \mat{\Lambda}_{Y}^{B} \\
    \mat{\Lambda}_{Z}^{B}
    \end{bmatrix} =  \begin{bmatrix}
    \mat{\Lambda}_{X}^{A} \\
    \mat{\Lambda}_{Y}^{A} \\
    \mat{\Lambda}_{Z}^{A}
    \end{bmatrix}\mat{R}_{3} , \label{eq:rotations}
\end{align}
for orthogonal matrices  $\mat{R}_{1}, \mat{R}_{2}$, and $\mat{R}_{3}$. The alignment of $\mat{\Lambda}_{X}^{A}$ and $\mat{\Lambda}_{X}^{B}$ is an orthogonal Procrustes problem. Let $\mat{R}$ be the solution to the optimisation problem
\begin{align}
    \mat{R} &= \argmin \ \lVert \mat{\Lambda}_{X}^{A} - \mat{\Lambda}_{X}^{B}\mat{R} \rVert_{F}, \quad \text{ subject to }  \mat{R}^{\T}\mat{R} = \mat{I}. \label{eq:procrustes_problem}
\end{align}
Assuming that  $\mat{\Lambda}_{X}^{A}$ and $\mat{\Lambda}_{X}^{B}$ are of full column rank, \citet{schonemann_1966_generalized} showed that there is a unique solution to \eqref{eq:procrustes_problem}. As $\text{rank}(\mat{\Lambda}_{X}^{A})=\text{rank}(\mat{\Lambda}_{X}^{B})=\text{rank}(\mat{\Lambda}_{X})$, both $\mat{\Lambda}_{X}^{A}$ and $\mat{\Lambda}_{X}^{B}$ are  of rank $q$ under Assumption \ref{assump:low}.  Define  $\mat{M} = (\mat{\Lambda}_{X}^{B})^{\T}{\mat{\Lambda}_{X}^{A}}$ and let the singular value decomposition of $\mat{M}$ be given by
    $\mat{M} =\mat{W}\mat{D}\mat{Q}^{\T} $. Then using the result from \citet{schonemann_1966_generalized},  the unique solution to \eqref{eq:procrustes_problem} is given by $\mat{R} = \mat{W}\mat{Q}^{\T}$. The uniqueness of the solution implies that $\mat{R}=\mat{R}_{3}^{\T}$ as $\Lambda_{X}^{B}\mat{R}_{3}\mat{R}_{3}^{\T} = \mat{\Lambda}_{X}^{A}$ from \eqref{eq:rotations}. Then $\mat{\Lambda}_{Z}^{B}\mat{R} = \mat{\Lambda}_{Z}^{B}\mat{R}_{3}^{\T} =\mat{\Lambda}_{Z}^{A}\mat{R}_{3}\mat{R}_{3}^{\T} = \mat{\Lambda}_{Z}^{A}$ again using \eqref{eq:rotations}. Finally, $\mat{\Lambda}_{Y}^{A}(\mat{\Lambda}_{Z}^{B}\mat{R})^{\T} = \mat{\Lambda}_{Y}^{A}(\mat{\Lambda}_{Z}^{A})^{\T}  = \mat{\Lambda}_{Y}\mat{R}_{1}\mat{R}_{1}^{\T}\mat{\Lambda}_{Z}^{\T} = \mat{\Lambda}_{Y}\mat{\Lambda}_{Z}^{\T}$.

\subsection{Proof of Theorem \ref{thm:identifiability}}
Using Theorem 5.1 in \cite{anderson_1956_statistical}, Assumption 2 guarantees that if
\begin{align*}
  \begin{pmatrix}
     \mat{\Lambda}_{X}\mat{\Lambda}_{X}^{\T} + \mat{\Psi}_{X} & \mat{\Lambda}_{X}\mat{\Lambda}_{Y}^{\T} \\
      \mat{\Lambda}_{Y}\mat{\Lambda}_{X}^{\T} & \mat{\Lambda}_{Y}\mat{\Lambda}_{Y}^{\T} + \mat{\Psi}_{Y}
    \end{pmatrix} &= \begin{pmatrix}
     \mat{\Lambda}_{X}^{*}\mat{\Lambda}_{X}^{*\T} + \mat{\Psi}_{X}^{*} & \mat{\Lambda}_{X}^{*}\mat{\Lambda}_{Y}^{*\T} \\
      \mat{\Lambda}_{Y}^{*}\mat{\Lambda}_{X}^{*\T} & \mat{\Lambda}_{Y}^{*}\mat{\Lambda}_{Y}^{*\T} + \mat{\Psi}_{Y}^{*}
    \end{pmatrix}, \\
    \begin{pmatrix}
     \mat{\Lambda}_{X}\mat{\Lambda}_{Z}^{\T} + \mat{\Psi}_{X} & \mat{\Lambda}_{X}\mat{\Lambda}_{Z}^{\T} \\
      \mat{\Lambda}_{Y}\mat{\Lambda}_{Z}^{\T} & \mat{\Lambda}_{Z}\mat{\Lambda}_{Z}^{\T} + \mat{\Psi}_{Z}
    \end{pmatrix} &=    \begin{pmatrix}
     \mat{\Lambda}_{X}^{*}\mat{\Lambda}_{Z}^{*\T} + \mat{\Psi}_{X}^{*} & \mat{\Lambda}_{X}^{*}\mat{\Lambda}_{Z}^{*\T} \\
      \mat{\Lambda}_{Y}^{*}\mat{\Lambda}_{Z}^{*\T} & \mat{\Lambda}_{Z}^{*}\mat{\Lambda}_{Z}^{*\T} + \mat{\Psi}_{Z}^{*}
    \end{pmatrix}.
\end{align*}
then the uniquenesses are equal, $\mat{\Psi}_{X}=\mat{\Psi}_{X}^{*}$, $\mat{\Psi}_{Y}=\mat{\Psi}_{Y}^{*}$, and $\mat{\Psi}_{Z}=\mat{\Psi}_{Z}^{*}$, implying
\begin{align}
  \begin{pmatrix}
     \mat{\Lambda}_{X}\mat{\Lambda}_{X}^{\T}  & \mat{\Lambda}_{X}\mat{\Lambda}_{Y}^{\T} \\
      \mat{\Lambda}_{Y}\mat{\Lambda}_{X}^{\T} & \mat{\Lambda}_{Y}\mat{\Lambda}_{Y}^{\T}
    \end{pmatrix} &= \begin{pmatrix}
     \mat{\Lambda}_{X}^{*}\mat{\Lambda}_{X}^{*\T} & \mat{\Lambda}_{X}^{*}\mat{\Lambda}_{Y}^{*\T} \\
      \mat{\Lambda}_{Y}^{*}\mat{\Lambda}_{X}^{*\T} & \mat{\Lambda}_{Y}^{*}\mat{\Lambda}_{Y}^{*\T} 
    \end{pmatrix},  \label{eq:obsA}\\
    \begin{pmatrix}
     \mat{\Lambda}_{X}\mat{\Lambda}_{X}^{\T}  & \mat{\Lambda}_{X}\mat{\Lambda}_{Z}^{\T} \\
      \mat{\Lambda}_{Z}\mat{\Lambda}_{X}^{\T} & \mat{\Lambda}_{Z}\mat{\Lambda}_{Z}^{\T} 
    \end{pmatrix} &=    \begin{pmatrix}
     \mat{\Lambda}_{X}^{*}\mat{\Lambda}_{X}^{*\T} & \mat{\Lambda}_{X}^{*}\mat{\Lambda}_{Z}^{*\T} \\
      \mat{\Lambda}_{Z}^{*}\mat{\Lambda}_{X}^{*\T} & \mat{\Lambda}_{Z}^{*}\mat{\Lambda}_{Z}^{*\T} 
    \end{pmatrix}. \label{eq:obsB}
\end{align}
Using Lemma \ref{lem:low_rank_completion}, $\Lambda_{Y}\Lambda_{Z}^{\T}$ can be uniquely recovered given the matrices on the left-hand side of \eqref{eq:obsA} and \eqref{eq:obsB}. Likewise, $\Lambda_{Y}^{*}\Lambda_{Z}^{*\T}$ can be uniquely recovered given the matrices on the right hand side of \eqref{eq:obsA} and \eqref{eq:obsB}. It remains to show that $\Lambda_{Y}\Lambda_{Z}^{\T} = \Lambda_{Y}^{*}\Lambda_{Z}^{*\T}$. To do so, define the eigendecompositions
\begin{align*}
    \mat{V}_{A}\mat{D}_{A}\mat{V}_{A}^{\T} &= \begin{pmatrix}
    \mat{\Lambda}_{X}\mat{\Lambda}_{X}^{\T} & \mat{\Lambda}_{X}\mat{\Lambda}_{Y}^{\T} \\
    \mat{\Lambda}_{Y}\mat{\Lambda}_{X}^{\T} & \mat{\Lambda}_{Y}\mat{\Lambda}_{Y}^{\T}
    \end{pmatrix}=\begin{pmatrix}
    \mat{\Lambda}_{X}^{*}\mat{\Lambda}_{X}^{*\T} & \mat{\Lambda}_{X}^{*}\mat{\Lambda}_{Y}^{*\T} \\
    \mat{\Lambda}_{Y}^{*}\mat{\Lambda}_{X}^{*\T} & \mat{\Lambda}_{Y}^{*}\mat{\Lambda}_{Y}^{*\T}
    \end{pmatrix}
    , \\
    \mat{V}_{B}\mat{D}_{B}\mat{V}_{B}^{\T} &= \begin{pmatrix}
    \mat{\Lambda}_{X}\mat{\Lambda}_{X}^{\T} & \mat{\Lambda}_{X}\mat{\Lambda}_{Z}^{\T} \\
    \mat{\Lambda}_{Z}\mat{\Lambda}_{X}^{\T} & \mat{\Lambda}_{Z}\mat{\Lambda}_{Z}^{\T}
    \end{pmatrix}=
    \begin{pmatrix}
    \mat{\Lambda}_{X}^{*}\mat{\Lambda}_{X}^{*\T} & \mat{\Lambda}_{X}^{*}\mat{\Lambda}_{Z}^{*\T} \\
    \mat{\Lambda}_{Z}^{*}\mat{\Lambda}_{X}^{*\T} & \mat{\Lambda}_{Z}^{*}\mat{\Lambda}_{Z}^{*\T}
    \end{pmatrix},
\end{align*}
and the rotated and scaled eigenvectors
\begin{align*}
    \begin{pmatrix}\mat{\mat{\Gamma}}_{X}^{A} \\ \mat{\mat{\Gamma}}_{Y}^{A} \end{pmatrix}&= \mat{V}_{A}\mat{D}_{A}^{1/2}, \quad 
   \begin{pmatrix} \mat{\mat{\Gamma}}_{X}^{B} \\ \mat{\mat{\Gamma}}_{Z}^{B}\end{pmatrix} = \mat{V}_{B}\mat{D}_{B}^{1/2}.
 \end{align*}
Using Assumption 1 and Lemma 1, the equality
\begin{align}
    \mat{\Lambda}_{Y}\mat{\Lambda}_{Z}^{\T} &= \mat{\Lambda}_{Y}^{*}\mat{\Lambda}_{Z}^{*\T}= \mat{\Gamma}_{Y}^{A}\mat{\Gamma}_{Z}^{B}\mat{W}\mat{Q}^{\T}, \label{eq:YZ_factors}
\end{align}
must hold, where $\mat{W}$ and $\mat{Q}$ are the left and right singular vectors of the matrix
     $\mat{M} = (\mat{\Gamma}_{X}^{B})^{\T}{\mat{\Gamma}_{X}^{A}} 
     = \mat{W}\mat{D}\mat{Q}^{\T}$. Combining the equalities in \eqref{eq:obsA}, \eqref{eq:obsB} and \eqref{eq:YZ_factors} gives the main result
     \begin{align*}
   \begin{pmatrix}
    \mat{\Lambda}_{X}\mat{\Lambda}_{X}^{\T} & \mat{\Lambda}_{X}\mat{\Lambda}_{Y}^{\T} & \mat{\Lambda}_{X}\mat{\Lambda}_{Z}^{\T} \\
    \mat{\Lambda}_{Y}\mat{\Lambda}_{X}^{\T} & \mat{\Lambda}_{Y}\mat{\Lambda}_{Y}^{\T} & \mat{\Lambda}_{Y}\mat{\Lambda}_{Z}^{\T} \\
    \mat{\Lambda}_{Z}\mat{\Lambda}_{X}^{\T} & \mat{\Lambda}_{Z}\mat{\Lambda}_{Y}^{\T} & \mat{\Lambda}_{Z}\mat{\Lambda}_{Z}^{\T}
    \end{pmatrix} &= 
   \begin{pmatrix}
    \mat{\Lambda}_{X}^{*}\mat{\Lambda}_{X}^{*\T} & \mat{\Lambda}_{X}^{*}\mat{\Lambda}_{Y}^{*\T} & \mat{\Lambda}_{X}^{*}\mat{\Lambda}_{Z}^{*\T} \\
    \mat{\Lambda}_{Y}^{*}\mat{\Lambda}_{X}^{*\T} & \mat{\Lambda}_{Y}^{*}\mat{\Lambda}_{Y}^{*\T} & \mat{\Lambda}_{Y}^{*}\mat{\Lambda}_{Z}^{*\T} \\
    \mat{\Lambda}_{Z}^{*}\mat{\Lambda}_{X}^{*\T} & \mat{\Lambda}_{Z}^{*}\mat{\Lambda}_{Y}^{*\T} & \mat{\Lambda}_{Z}^{*}\mat{\Lambda}_{Z}^{*\T}
    \end{pmatrix}.
\end{align*}

\end{document}